# On A Testing and Implementation of Quantum Gate and Measurement Emulator (QGAME)

A.B. Mutiara [1], R.Refianti [2], J.S.K. Karamoy [3]

*Faculty of Computer Science and Information Technology, Gunadarma University*
*Jl. Margonda Raya No.100, Depok 16424, West Java, Indonesia*
[1] amutiara@staff.gunadarma.ac.id
[2] rina@staff.gunadarma.ac.id
[3] jskkaramoy@gmail.com

*Abstract*—**Today, people are looking forward to get an awesome computational power. This kind of desire can be answered by quantum computing. By adopting quantum mechanics theory, it can generate a very fast computation result. As known, quantum mechanics can establish that particle can also become wave; it shows that electron can be in duality. Through this theory, even a human teleportation is issued can be really happened in the future. However, it needs a high requirement of hardware support to implement the real quantum computing. That is why it is difficult to bring quantum computing into reality. This research presents a study about quantum computing. Here it is studied, a specialty of quantum computing, like superposition, as if the classical computer can do it. Since there was a marvellous research about quantum computer simulation that runs on classical computer, this research provides an analysis about our testing and implementation of Quantum Gate and Measurement Emulator (QGAME). Our analysis, testing and implementation are based on a method that always use in the software engineering field.**
 Keyword-Quantum Computation, Quantum Computer Simulation, Software Engineering

## I. INTRODUCTION

An awesome computational power will be presented to humanity in time quantum computing is fully realized. Unfortunately, it is hard to fully comprehend what physic phenomenon quantum computing deals with. Quantum computing adopts quantum mechanics theory sparked by Max Planck that has some impressive and arduous phenomena to process information. The quantum theory was studied more by Albert Einstein and other scientists. It is such a very long term research which is hard to meet the finish line. As a famous physicist once said, Richard Feynman, "If you think you understand quantum mechanics, you don't understand quantum mechanics."

Quantum mechanics bring up the fact that an electron can be a particle and a wave. Particle is a principle unit of thing or energy which can be stated simply as a small localized object that forms several physical properties, whereas wave is the energy. How can particle become waves? Amazingly, it is possible in quantum mechanics. Event like this is called wave-particle duality. To describe the truth of this quantum mechanics theory, Austrian physicist Erwin Schrödinger delivered experiment known as Schrödinger's cat [5]. According to this experiment, some properties of particles were not decided until an outsider forces them to choose by measuring them. Schrödinger envisioned a cat inside a box that contained a small amount of a radioactive substance. Within an hour, there would be a 50 percent chance the substance would decay, releasing poison into the box, and a 50 percent chance the substance would not decay and no poison would be released. According to the rules of quantum mechanics, the cat was neither dead nor alive until the box was opened and an outside observer "measured" the situation. While the box remained closed, the whole system was suspended in a state of uncertainty where the cat was both dead and alive.

To make use of quantum mechanical effect in computation, a quantum computer is needed. Quantum computers are different from digital computers based on transistors. Whereas digital computers require data to be encoded into binary digits (bits), quantum computation uses quantum properties to represent data and perform operations on these data [1]. However, a quantum computer of significant size is difficult to build. No one has yet succeeded in such research. There are fundamental difficulties will be met for a large-scale quantum computer being built. Therefore, the problem of this research is how to simulate quantum computation on classical computer.

This research presents study of a quantum computer simulation proposed by Lee Spector [7]. Spector created a simulator in such a way that user could run quantum computing algorithm on classical computer. Therefore, the objectives of this research are to provide development process of quantum computing simulator and to analyse the program. This research can be used to comprehend quantum programming approach. It can be used as reference to build a quantum programming emulator.

## II. Quantum Computing

Quantum computing is computation processes those quantum mechanical properties of information-processing hardware impact on. It allows doing things that classical computing cannot. The most spectacular known quantum computing in 1994 was the complexity proposed by Peter Shor's algorithm for factoring large number [6]. If classical algorithms required about 80 billion to factor 5.000 digit numbers, by contrast the quantum algorithms only required less than two seconds. There was also modest quantum algorithm that was proposed by Lov Grover. Grover's quantum search algorithm achieved a quadratic speedup over the best classical algorithms for finding a single "marked" item in a database [2]. If classical algorithms had to test half of the n items in the database, the quantum algorithm was able to find the item after marking only about √n queries.

Quantum computing also provides benefits beside two algorithms presented above. For example, quantum states are "tamper resistant" in a certain sense, and this property can be leveraged to provide secure communication channels upon which it is theoretically impossible to eavesdrop. Some of the schemes for such channels require relatively little in the way of quantum hardware engineering, and quantum information technology products for secure communications are already commercially available [7].

In classical computing, the fundamental unit of information is the bit, which can exist in one of two states (conventionally labelled "0" and "1") [7]. A variable or computed quantity that is only able to have two possible values can be a definition for bit in computing. Denoted by the numerical digits 0 and 1, interpreted as binary digits, interpreted as logical values (true/false) are kind of characteristic of bit. In quantum computing the fundamental unit of information is the qubit (quantum bit), which can also exist in one of two "computational basis" states (conventionally labelled using Paul Dirac's "bra-ket" notation as |0> and |1>). But unlike the bit, the qubit can also exist in a superposition of |0> and |1> represented as $\alpha_0|0> + \alpha_1|1>$, where $|\alpha_0|^2 + |\alpha_1|^2 = 1$. Within a bit the information is stored either as a 0 or 1. In contrast to a bit, a qubit stores information as 0, 1, and the superposition of all numbers between 0 and 1. As an analogy: if a bit is like a light switch - either on or off - a qubit is like a dimmer switch, but one which is set at all positions simultaneously [8]. To be clear about difference between bit and qubit can be described in Figure 1. The description shows us if bit has a exact value either 0 or 1, qubit has many possibilities of values described like a ball.

Qubits in a quantum computer has a special characteristic called by entanglement. Qubits can become "entangled" with one another, and this entanglement underlies several interesting quantum algorithms [7, 3]. It clearly distinguishes qubit and classical bit. Entanglement also allows multiple states to be acted on simultaneously, unlike classical bit that is only able to have one value at a time. Entanglement is a necessary ingredient of any quantum computation that cannot be done efficiently on a classical computer. Entanglement is also said to be "spooky action at a distance" in that when you collapse the wave function of two entangled states they take opposite values and this change is somehow shared instantaneously across distances faster than the speed of light. A number of entangled qubits taken together is a qubit register [4].

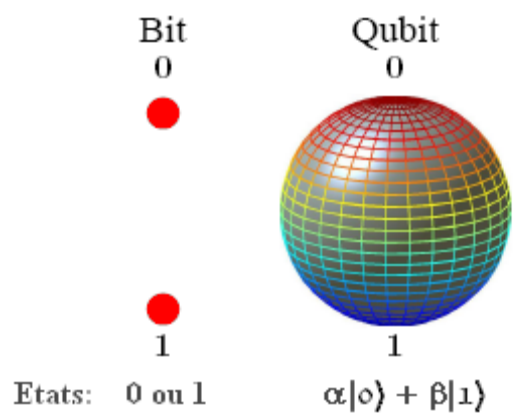

Fig.1. The Difference between Bit and Qubit [4]

The basic gate of classical computer are the two-bit gates, AND and OR, and the one-bit gate NOT. Gate functions in classical logic are often represented using truth tables. Quantum computation can be viewed, mathematically, as a generalization of the classical matrix model. The gates are QNOT, U-Theta, Square Root of NOT, HADAMARD, U2, Controlled Phase, SWAP gate and many more. HADAMARD gate plays role to transform qubit into superposition state.

## III. QUANTUM GATE AND MEASUREMENT EMULATOR (QGAME)

Theoretically, quantum simulation is able to be developed with full matrix mathematics, since the fundamental requirement to simulate was ability to simulate the execution of arbitrary sequences of quantum gates [7]. To apply the concept above, it is impossible to allow quantum gates described in the previous section for the simulation system because of the inappropriate size. How for instance the 2x2 QNOT matrixes does implement to 3-qubit system? The matrix needs to be expanded.

QGAME was based on the "quantum gate array" model of quantum computation, in which quantum "gates" (represented as square matrices) were applied to a register of qubits (via tensor product formation and matrix multiplication). QGAME always starts with all qubits having the value zero (in the state $|00...0\rangle$), applies a sequence of gates, and returns values about the resulting state. Measurement gates caused the system to branch, following one execution path (with the appropriate quantum state collapse) for each possible value. Final measurements were made across the end-states of all of the resulting branches [7].

## IV. SOFTWARE ENGINEERING ANALYSIS

### A. User Requirements

The software has to recognize syntax of quantum programs and provide interpreter that simulates the quantum computing execution. Actually, the basic requirements are fully covered by the proposed QGAME++. It also provides a way to specify algorithms that include calls to "oracle" gates with any number of inputs and one output. The program also support if one wants to run the program with various instances of the oracle and collect statistics over all of the results. QGAME is able to simulate the effects of single-qubit measurements during the execution of a quantum program, and allows for the outcomes of those measurements to influence the remainder of the simulation. The program is also able to limit the number of times that an oracle can be called during a single simulation. The benefit is the program allows the inclusion of gates with arbitrary unitary matrices. In order to apply QGAME++, there must be a quantum program. However, these are the quantum program functions that will be recognized by the system. There are two kinds of functions, gate and instruction (see Table I).

TABLE I
Quantum Program Syntax

| Syntax | Description |
|---|---|
| (QNOT q) | Applies a QNOT gate to specified qubit q |
| (CNOT q) | Applies a quantum controlled NOT gate to the specified control and target qubits q |
| (SRN q) | Applies a quantum square root of NOT gate to the specified control and target qubits q |
| (HADAMARD q) | Applies a quantum hadamard gate to the specified qubits q |
| (U-THETA q $\theta$) | Applies a quantum u-theta gate to the specified qubits q |
| (U2 q $\phi$ $\theta$ $\psi$ $\alpha$) | Applies a quantum U2 gate to the specified qubits q |
| (CPHASE $q_{control}$ $q_{target}$ $\alpha$) | Applies a quantum u-theta gate to the specified control and target qubits q |
| (SWAP $q_{control}$ $q_{target}$) | Applies a quantum swap gate to the specified control and target qubits q |
| (ORACLE $\Omega$ $q_1$ $q_2$ … $q_n$ $q_{out}$) | Works just like the previous ORACLE function with the specified times of execution |
| (MEASURED q) ...branch1... (END) ...branch0... (END) | The syntax for a QGAME measurement |
| (HALT) | Terminates the current simulation |
| (PRINTAMPS) | Prints the amplitudes of the executing quantum system |

The software needs a screen to enter the program arguments such as the number of qubits the system should use for executing the program and to display the results of simulation and measurement. QGame++ should run on any system supporting the C++ standard based on the Linux Operating System. Technically, this software was made to produce a user friendly interface for user to use the simulator well, produce the desired system output of quantum system result, be able to run several times on different test cases, and produce some statistical data.

### B. Functional Point Analysis

This is analysis about program features. Here is explained, output and input the program should generate. Files and other parameter the program should have also are explained here.

1. External Outputs: The program will produce output(s) as follows: 1) the number of "misses"; cases in which the measured value will, with probability greater than the specified threshold, fail to equal the desired output 2) the maximum probability of error for any provided case 3) the average probability of error for all provided cases 4) the maximum number of expected oracle calls across all cases 5) the number of expected oracle calls averaged across all cases 6) in other execution, provided also the measurement history [7].

2. External Inputs: There are fives input features provided by the application: 1) the program, that is the program to be tested, in QGAME program syntax 2) the number of qubits, that is the number of qubits to be simulated in the quantum computer 3) the cases, that is a parenthesized list of "(oracle-truth-table output)" pairs, where each oracle-truth-table is a parenthesized list of 0s and 1s specifying the right-hand (output) column of the oracle's truth table (where the rows are listed in binary order), and where the output is the correct non-negative integer answer for the given truth table; the test compares this number to the number read from the final measurement qubits at the end of the computation 4) the final measurement qubits, that is a parenthesized list of indices specifying the qubits upon which final measurements will be performed, with the most significant qubit listed first and the least significant qubit listed last and 5) the threshold, that is the probability of error below which a run is considered successful for the sake of the "misses" component of the return value. This is typically set to something like 0.48, which is usually far enough from 0.5 to ensure that the "better than random guessing" performance of the algorithm is not due to accumulated round-o errors [7].

3. External Interface Files: The application refers to library qgame++ that will be used by the simulation program. It also needs an input le such as quantum_program_1.qcp, the input le for the system simulates.

4. External Inquiry: The application does not have such the external output without any calculation as the external inquiry.

5. Internal Logical Files: This application does not have such the database as the internal logical files.

C. Architecture Design

1) *General Description:* The basic principle of the system is to represent the emulating result to the user that is appropriate with the instruction given by user. There will be two different processes done by the system. In general, Fig. 2 describes how the system works.

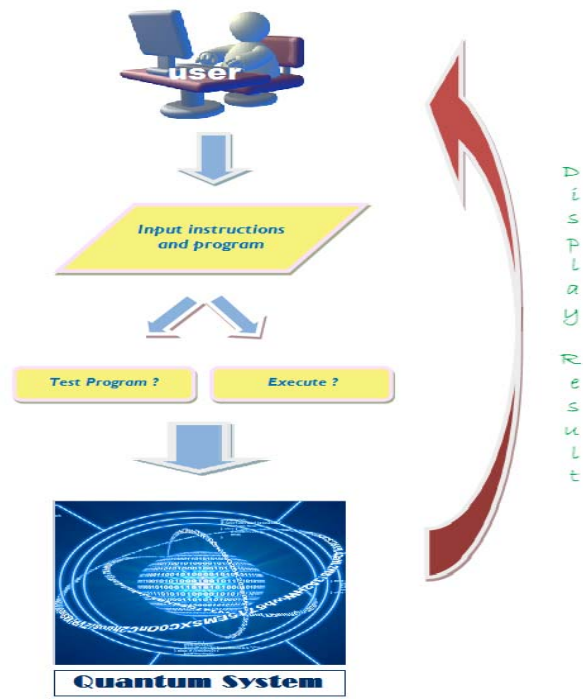

Fig.2. General Description of the System

There is an interface provided for user to prompt some instructions based on the system rule. There will be some parameters related to the prompted instructions that will be read by system in order to answer the user needs. First of all, user prompts some instructions to the system. The input will then be extracted by system. Of course, the role of this work is done by a quantum system. This quantum system will check the completeness of the prompted instructions. If there is one or more instruction missing, the system will ask user to prompt the remaining input. After all input is generated well, the system will check the instruction type. There are two instructions types as the main feature of the application, which are test and execute. The difference of that twos is already explained in previous sections. At the end of process, the type "test" will result a test result, whereas the type "execute" will result a measurement history. All results will be sent to the user interface system.

2) *Use-Case Diagram:* Presented here a use-case diagram to describe who will use the system and how the user interacts with the system. Based on Fig. 3, user is able run le, see simulation test result, execute file, see measurement history, see help, and see version. As the application runs, the information of application usage and the example are displayed, so user is able to simulate immediately. Actually the system has subsystem which is program simulation test and program execution.

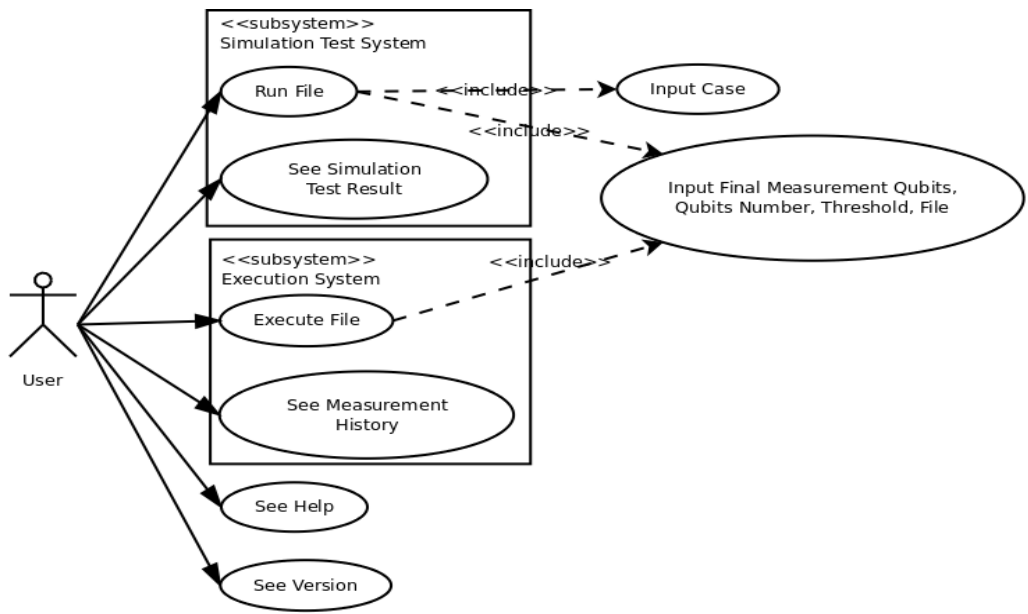

Fig.3. Use-Case Diagram

To run program test, user has to input list of test cases, qubits number, final measurement qubits, threshold, and filename of the tested program. List of test cases consists of (the right side of) an oracle truth table and the desired results one wants to measure. Qubits number specifies the number of qubits, final measurement qubits specify the qubits of final measurement, threshold specify threshold for error to be compared with the probability of error, and file specifies the program listing.

The system cannot run perfectly if one of the inputs above is not available. If user only input the cases, then program will ask user to input the other. As the inputs entered to the system, the application will generate the statistical data test as the result. The use-case of this description is simulation test result. The result contains of information about probability of desired output below threshold, maximum error that occurred during test, the average error, maximum number of expected oracle calls and average expected oracles. It will also display the inputs by user.

The duty of input deals just the same with use-case execute file. User has also to input the final measurement qubits, qubits number, threshold, and le, but the case is not needed. The program specified by filename input will be run and return the list of resulting system. User is then able to see measurement history on the screen. Beside run or test quantum program and execute it, user is able to see help feature of the application. It contains information of usage syntax and example of the usage. User is also able to see version of the application.

3) *Class Diagram:* Based on use-case diagram, system needs to have classes that handle quantum simulation system, measurement, result, unexpected error, and etc. The class will specialise each object characteristic. Attributes and operations of each classed will also be described here. The class diagram of the system can be seen in Fig. 4.

4) *Sequence Diagram "Test Program":* The sequence diagram of quantum program simulation test is described in Figure 5. User inputs cases, qubits number, nal measurement qubits, threshold, and program le to the system that will immediately managed by Qsys. Qsys will process all inputs by user by calling class that will handle each input based on the purpose of the class. The result will be displayed to user as Result structure consist of misses, maxError, avgError, maxExpOracles, and avgExpOracles.

5) *Sequence Diagram "Execute":* The sequence diagram of quantum program execution is described in Figure 6. User inputs qubits number, final measurement qubits, threshold, and program file to the system that will immediately managed by Qsys. Qsys will process all inputs by user by calling class that will handle each input based on the purpose of the class. The result will be displayed to user as measurement history.

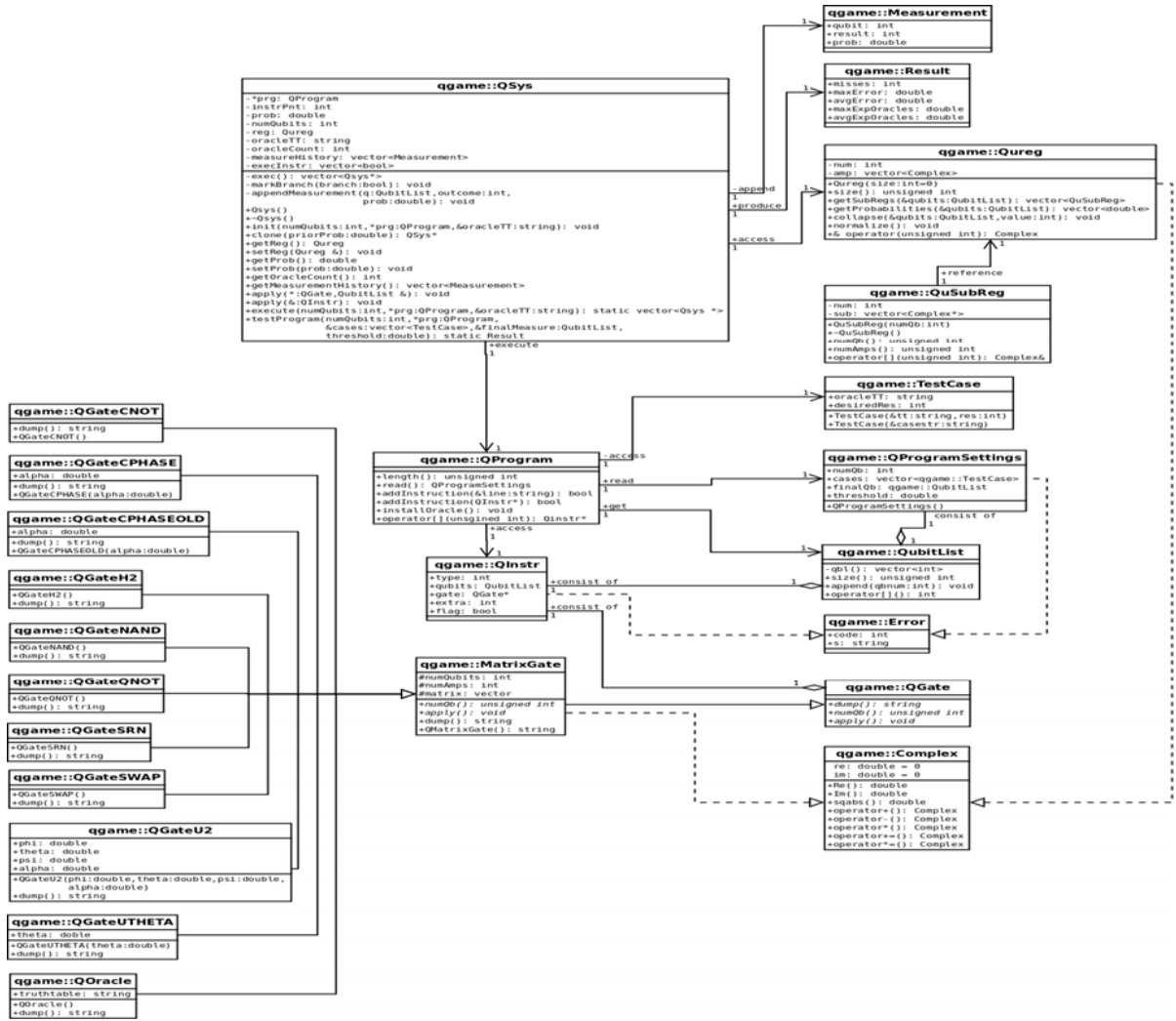

Fig.4. Class Diagram

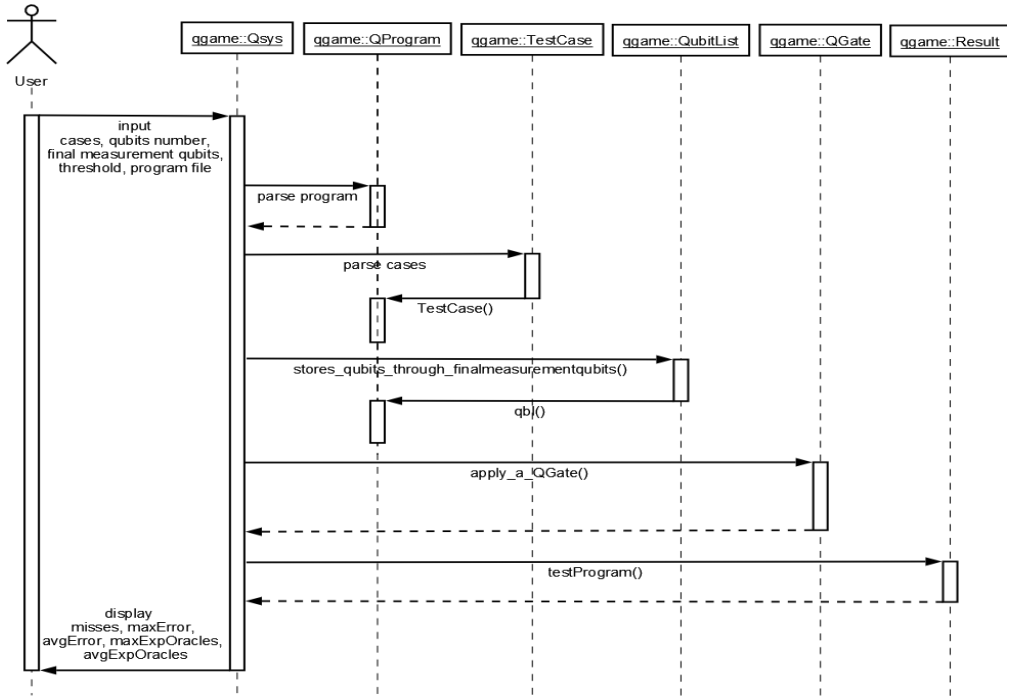

Fig.5. Sequence Diagram "Test Program"

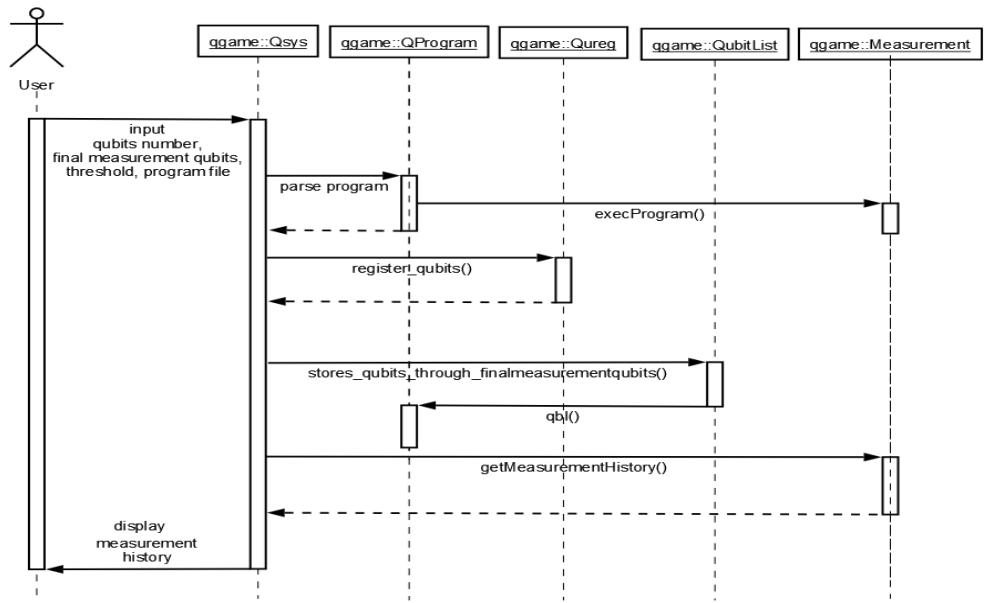

Fig.6. Sequence Diagram "Execute"

*D. Testing*

*1) Feature Testing:* Based on the purpose of this research, the application system has to be able to read instructions of given quantum program les. The application system will throws error exception if it reads unknown format of given instructions in the program. Besides reading quantum program, it has to be able to read program arguments. At the beginning, user will be displayed an interface to input some arguments to run the application. In this stage, the application system plays role for recognizing given program arguments by user based on the argument rule. If the argument prompted by user is not appropriate, the application system will throw error exception. The other feature tested on the application is formatting values for output as explained above. If the application system returns an incorrect value, it will debug the resulted output. Table II described test cases and the results.

TABLE II
Feature Testing

| Test Unit | Test Case | Result |
|---|---|---|
| Read Program | quantum_program_1.qcp input file (using qcp extension) (HADAMARD 2) | Successful! Program is read and printed |
| Read Program | quantum_program_2.txt input file (using qcp extension) (HADAMARD 2) | Successful! Program is read and printed |
| Read Program | quantum_program_3.qcp (HADAMARD 2 | <EMPTY> |
| Read Program | quantum_program_4.qcp HADAMARD 2 | <EMPTY> |
| Read Program | quantum_program_5.qcp (HADAMARD 2) (QNOT 2) | Successful! Program is read and printed |
| Read Program | quantum_program_6.qcp (HADAMARD 2) (QNOT 2) | Successful! Program is read and printed |
| Read Program | quantum_program_7.qcp (HADAMARD 2) (CNOT 2) (HADAMARD 1) | Syntax error Program aborted. |
| Format Values for Output | (HADAMARD -2) | Warning: Non-unitary transformation p=0 |
| Read Program Arguments | qgame c 0000-0 0001-1 | Usage: qgame [OPTIONS] FILE ... |
| Read Program Arguments | qgame -c 0000-0 0001-1 | I/O Error Program aborted. |
| Read Program Arguments | qgame -c 0000-0 0001-1 -n 3 -f 0 1 1 -t 1 quantum_program_1.qcp | Successful! Test is run. |
| Read Program Arguments | qgame -v | Successful! Display version |

2) *Functional Testing:* Both test program and execute function were tested. In this phase, execute result is analyzed. Here are conducted 4 cases of tests which are based on the oracle truth tables. The example of instructions given to the system is:

qgame −x 1000 −n 3 −f 1 2 −t 1 test1.qcp

Inputs:

Oracle Truth Table: The quantum program will be executed using oracle truth table 1000 which is initialized by -x. There are 4 kind of execution in this trial and the result is described in TABLE III.

Number of Qubits: The system uses 3-qubit as initialized by -n, so the quan-tum gate will be applied to the program is 3-qubit gate.

Final measurement qubits: The system uses two nal measurement qubits. The most signi cant qubit is listed last, so qubit 2 will be referred as the most significant one.

Threshold: The system uses 1 as threshold value, means that the system only allows probability of error equal or less than 1.

Quantum program: The system uses test1.qcp as the quantum program be-ing tested. test1.qcp is the grover database search program that is already explained above.

Outputs:

The explanation of the output is as follows. The output shows user the mea-surement history of grover database search program execution. It asks system to

TABLE III.
Functional Testing - Execute

**1000**

| Register | Amplitude | Probability |
|---|---|---|
| $|000\rangle$ | 0.707 | 0.5 |
| $|001\rangle$ | -0.707 | 0.5 |
| $|010\rangle$ | $2.36 \times 10^{-17}$ | $5.60 \times 10^{-34}$ |
| $|011\rangle$ | $-6.29 \times 10^{-17}$ | $3.96 \times 10^{-34}$ |
| $|100\rangle$ | $4.33 \times 10^{-17}$ | $1.87 \times 10^{-33}$ |
| $|101\rangle$ | $-9.8 \times 10^{-17}$ | $5.60 \times 10^{-34}$ |
| $|110\rangle$ | $1.96 \times 10^{-17}$ | $3.85 \times 10^{-34}$ |
| $|111\rangle$ | $1.96 \times 10^{-17}$ | $3.85 \times 10^{-34}$ |

(1)

**0100**

| Register | Amplitude | Probability |
|---|---|---|
| $|000\rangle$ | $6.29 \times 10^{-17}$ | $3.96 \times 10^{-34}$ |
| $|001\rangle$ | $-2.36 \times 10^{-17}$ | $5.60 \times 10^{-34}$ |
| $|010\rangle$ | -0.707 | 0.5 |
| $|011\rangle$ | 0.707 | 0.5 |
| $|100\rangle$ | $-1.96 \times 10^{-17}$ | $3.85 \times 10^{-34}$ |
| $|101\rangle$ | $-1.96 \times 10^{-17}$ | $3.85 \times 10^{-34}$ |
| $|110\rangle$ | $-4.33 \times 10^{-17}$ | $1.87 \times 10^{-33}$ |
| $|111\rangle$ | $9.8 \times 10^{-17}$ | $5.60 \times 10^{-34}$ |

(2)

**0010**

| Register | Amplitude | Probability |
|---|---|---|
| $|000\rangle$ | $4.33 \times 10^{-17}$ | $1.87 \times 10^{-33}$ |
| $|001\rangle$ | $1.22 \times 10^{-17}$ | $1.49 \times 10^{-34}$ |
| $|010\rangle$ | $-1.96 \times 10^{-17}$ | $3.85 \times 10^{-34}$ |
| $|011\rangle$ | $-1.96 \times 10^{-17}$ | $3.85 \times 10^{-34}$ |
| $|100\rangle$ | -0.707 | 0.5 |
| $|101\rangle$ | 0.707 | 0.5 |
| $|110\rangle$ | $-2.36 \times 10^{-17}$ | $5.60 \times 10^{-34}$ |
| $|111\rangle$ | $6.29 \times 10^{-17}$ | $3.96 \times 10^{-34}$ |

(3)

**0001**

| Register | Amplitude | Probability |
|---|---|---|
| $|000\rangle$ | $1.96 \times 10^{-17}$ | $3.85 \times 10^{-34}$ |
| $|001\rangle$ | $1.96 \times 10^{-17}$ | $3.85 \times 10^{-34}$ |
| $|010\rangle$ | $-4.33 \times 10^{-17}$ | $1.87 \times 10^{-33}$ |
| $|011\rangle$ | $-1.22 \times 10^{-17}$ | $1.49 \times 10^{-34}$ |
| $|100\rangle$ | $-6.29 \times 10^{-17}$ | $3.96 \times 10^{-34}$ |
| $|101\rangle$ | $2.36 \times 10^{-17}$ | $5.60 \times 10^{-34}$ |
| $|110\rangle$ | 0.707 | 0.5 |
| $|111\rangle$ | -0.707 | 0.5 |

(4)

find item in one of the quantum register. And the execution conducted here implements the probability that measurement of the qubit will find the corresponding state. As known, the superposition of these 3-qubit states is structured as

$\alpha_0|000\rangle + \alpha_1|001\rangle + \alpha_2|010\rangle + \alpha_3|011\rangle + \alpha_4|100\rangle + \alpha_5|101\rangle + \alpha_6|110\rangle + \alpha_7|111\rangle$,

where the amplitude of each qubit is as follow.

$\alpha_0 = 0.707$, with probability $= |\alpha_0|^2 = 0.5$

$\alpha_1 = -0.707$, with probability $= |\alpha_1|^2 = 0.5$

$\alpha_2 = 2.36 \times 10^{-17}$, with probability $= |\alpha_2|^2 = 5.6 \times 10^{-34}$

$\alpha_3 = -6.29 \times 10^{-17}$, with probability $= |\alpha_3|^2 = 3.95 \times 10^{-33}$

$\alpha_4 = 4.32 \times 10^{-17}$, with probability $= |\alpha_4|^2 = 1.87 \times 10^{-33}$

$\alpha_5 = -9.8 \times 10^{-17}$, with probability $= |\alpha_5|^2 = 9.76 \times 10^{-33}$

$\alpha_6 = 1.96 \times 10^{-17}$, with probability $= |\alpha_6|^2 = 3.85 \times 10^{-34}$

$\alpha_7 = 1.96 \times 10^{-17}$, with probability $= |\alpha_7|^2 = 3.85 \times 10^{-34}$

Based on the data above, the greatest probability of measured amplitudes is 0.5 which is in register $\alpha_0$ and $\alpha_1$. It means that the measurement of the qubit will nd the corresponding state in the $\alpha_0$ and $\alpha_1$ states for the oracle truth table 1000. The probability of the qubit will nd the corresponding state in the rest of 6 states is too small, up to $5.6 \times 10^{-34}$. It is quite impossible to find the item in $\alpha_2$ for example. However, it is only valid if the oracle truth table is 1000. The other executions have their own most possible corresponding state.

Based on the example above, it can be concluded that the state which the search item located can be found if user gives any oracle truth table to be measured. As it is available, system will calculate the amplitudes based on the algorithm. So, quantum computing is really related to measurement.

## V. SUMMARY AND CONCLUDING REMARKS

The development process of quantum computing simulator has already presented. The application that is proposed by Lee Spector written by Manual Manuel Nickschas is also tested. Based on the testing result, the application proves the "quantum" side of the computation and it seems to tell that quantum computing is really related to measurement. It can be seen by noticing one of quantum property which is superposition state. If there is n-state, one can determine the most corresponding state of n-state by measuring the probability of each state.

It is successful to simulate a quantum program with the significant result of the probability measured. The simulated quantum system is able to show user the fast computation. It proves that quantum computer is really powerful to answer user needs. The system is powerful, but it needs a significant enhancement. It is better if someday it is improved by transforming it into GUI application, so that user is easier to interact with the system. In addition, the visualization of the quantum process needs to be embedded to the system to enrich user comprehension.

ACKNOWLEDGMENT

A.B.M. and R.R. gratefully acknowledges a financial support of the Gunadarma Foundations.